\documentclass[preprint2]{aastex}

\slugcomment{}

\def\kms{km\thinspace s$^{-1}$}
\def\cm2{cm$^{-2}$}

\def\sun{_\odot}
 
\shortauthors{Taylor \& Wilson}
\shorttitle{Atomic Carbon in NGC\,604-2}

\begin{document}

\title{The Spatial Distribution of Atomic Carbon Emission in the
Giant Molecular Cloud NGC\,604-2}

\author{Christopher L. Taylor}
\vspace{0.2pt}
\affil{Ruhr-Universit\"at Bochum, Astronomisches Institut \\
Universit\"atsstr 150, D-44870 Bochum, Germany \\
and \\
Five College Radio Astronomy Observatory \\ University of Massachusetts \\
Amherst, MA, 01003 }

\authoraddr{Ruhr-Universit\"at Bochum, Astronomisches Insitut, 
Universit\"atsstr 150, D-44870 Bochum
Germany \linebreak{\it chris@fcrao1.astro.umass.edu}}

\author{Christine D. Wilson}

\vspace{0.2pt}
\affil{McMaster University \\ Department of Physics and Astronomy \\
Hamilton, Ontario, Canada, L8S 4M1}

\authoraddr{ \\ McMaster University \\ Department of Physics and Astronomy \\
Hamilton, Ontario, Canada, L8S 4M1
\linebreak {\it wilson@physics.mcmaster.ca}}

\vspace{.75in}


\begin{abstract}
We have mapped a giant molecular cloud in the giant HII region NGC\,604
in M33 in the 492 GHz $^3P_1 \rightarrow ^3P_0$ transition of neutral
atomic carbon using the James Clerk Maxwell Telescope.  We find the 
distribution of the [CI] emission to be asymmetric with respect to the
CO J=$1\to0$ emission, with the peak of the [CI] emission offset towards 
the direction of the center of the HII region.  In addition, the line
ratio I$_{[CI]}$/I$_{CO}$ is highest ($\sim$ 0.2) facing the HII region
and lowest ($\lesssim$ 0.1) away from it. These asymmetries indicate
an edge-on morphology where the [CI] emission is strongest on the side of 
the cloud facing the center of the HII region, and not detected at all on 
the opposite side  This suggests that the sources of the incident flux 
creating C from the dissociation of CO are the massive stars of the HII 
region.  The lowest line ratios are similar to what is observed in 
Galactic molecular clouds,  while the highest are similar to starburst 
galaxies and other regions of intense star formation.  The column density 
ratio, N(C)/N(H$_2$) is a few $\times$~10$^{-6}$, in general agreement with 
models of photodissociation regions.
\end{abstract}


\section{Introduction}

The fine structure line of atomic carbon at 492 GHz (607 $\mu$m) is 
one of the important cooling lines of photon dominated regions (PDRs).  
PDRs form the transition zones between the molecular interstellar 
medium (ISM) and the atomic ISM, where molecular gas is dissociated 
by ultraviolet photons from nearby star forming regions.  Maps of the 
distribution of [CI] emission in molecular clouds and on larger scales 
in nearby galaxies have revealed that it is often cospatial with the 
emission from CO (e.g., Keene et al. 1985; Israel, White 
\& Baas 1995), despite the fact that the ionization 
energy of carbon is very close to the photodissociation energy of CO.  
From these energy considerations, [CI] would be expected to exist only 
in a thin layer at the surface of molecular clouds.  The similarity 
between the spatial distributions of CO and [CI] is evidence for a 
clumpy molecular ISM that allows ultraviolet photons to penetrate into 
the interiors of molecular clouds (White \& Padman 1991). 

The ratio of atomic carbon to CO, N(C)/N(CO), depends on which chemical
reactions occur in the PDR, which in turn depends upon the
degree of ionization of the gas.  When the ionization fraction exceeds
a critical value, the gas phase chemistry is dominated by charge transfer 
reactions with H$^+$. These reactions suppress the amount of H$_3^+$, 
the most important species for reactions that destroy atomic 
carbon. Thus, at high ionization fractions, the abundance of C can remain 
high, while at lower ionization fractions H$_3^+$ reactions dominate 
and C is destroyed (Graedel, Langer \& Frerking et al. 1982; 
Flower et al. 1994).  Both UV photons and cosmic rays can influence
the ionization fraction.  A strong UV field incident upon a molecular 
cloud will eventually be stopped by the increasing optical depth as the 
radiation penetrates the cloud.  
Thus, the distribution of [CI] emission should appear asymmetric.  The 
flux of energetic cosmic rays, on the other hand, will be isotropic 
within a galaxy and the cosmic rays will be capable of penetrating
the entire molecular cloud.  If the cosmic rays are a more important
source of ionization than the UV flux, the [CI] distribution should 
appear similar to the CO distribution, with both species tracing the 
overall distribution of carbon in the ISM.

We present observations of the [CI] line at 492 GHz of the giant molecular
cloud (GMC) NGC\,604-2, in the giant HII region NGC\,604, which is in the Local 
Group galaxy M33.  By observing a GMC in a nearby galaxy (D = 
0.84 Mpc; Freedman, Wilson, \& Madore 1991), we can map the extent of 
the cloud in a comparatively short time, and thus determine the spatial 
distribution of the [CI] emission.  NGC\,604-2 has a molecular mass of 
$6.3~\times~10^5$ M$\sun$, a diameter of $\sim$ 32 pc, and a linewidth of 
$\sim$ 11 \kms\thinspace (Wilson \& Scoville 1992).  It is similar to the larger 
GMCs in the Milky Way Galaxy. NGC\,604 is the most luminous HII region in 
M33, and is the second nearest giant HII region, the nearest being 30 Doradus 
in the Large Magellanic Cloud.  Although these two HII regions are comparable 
in size, the H$\alpha$ luminosity of NGC\,604 is approximately half that of 
30 Doradus (Kennicutt 1984).  The most 
important difference between the two regions is the distribution of their 
ionizing stars: 30 Doradus is dominated by the R136 cluster (Walborn 1991), 
while in NGC\,604 the distribution of massive stars is not so concentrated 
towards the center of the HII region (Drissen et al. 1993).  Hunter et al. 
(1996) have calculated the average density of O stars in NGC\, 604 to be 
0.0018 stars pc$^{-2}$, 100 times lower than in R136 and more than a factor 
of 2 lower than the average density over the entire 30 Doradus region. 

\section{Observations and Data Reduction}

We observed the GMC NGC\,604-2 in the $^3P_1 \rightarrow 
^3P_0$ transition of [CI] at 492.2 GHz using the James Clerk Maxwell 
Telescope (JCMT) on Mauna Kea during 1997 November 21 - 22,  1997 December 
12, 1998 December 27 and 1999 July 28.  The half power beam width at this 
frequency is 12\arcsec, so a map was constructed using a grid with 
5\arcsec\ spacing.  The axes of this grid are rotated 45$^{\circ}$\ with 
respect to the equatorial coordinate system.  Calibration observations 
of W75N and W3(OH) differed from the standard JCMT reference spectra by 20\% 
on 21 November, and by 5\% on the other dates, so we adopt as the uncertainty 
in the absolute calibration a factor of 20\%.  These reference spectra
are available on the \anchor{http://www.jach.hawaii.edu/JACpublic/JCMT}
{JCMT web page: http://www.jach.hawaii.edu/JACpublic/JCMT}. 

The pointing was checked every 1 to 2 hours by observing the planets Mars, 
Jupiter, and Uranus, and the source NGC~7538 IRS 1.  The pointing of the 
telescope drifted during the observations, requiring us to correct for 
this drift.  Immediately after a pointing check the pointing accuracy
was 1\arcsec\ or better, small compared to the beam size.  After 1 or
2 hours, the drifting resulted in offsets relative to
the intended position that were a significant fraction of the beam.
On 1997 November 21, the maximum drift rate measured was 
1\arcsec\ per hour, while on November 22 the drift reached 1.5\arcsec\ per 
hour, and on December 12 1.7\arcsec\ per hour.  For observations taken 
more than an hour after a pointing check, a correction was applied in the
data reduction stage.  The altitude-azimuth coordinate system in which 
pointing is checked rotates with respect to equatorial coordinates through 
the night, so a systematic drift in alt-az coordinates does 
not correspond to a constant offset in position in equatorial coordinates.  
A log of the observed positions, the applied pointing corrections in 
equatorial coordinates, and the rms noise for the final spectra is given 
in Table~1.

\tablenum{1}
\placetable{table1}
\label{table1}

The data reduction was carried out using the SPECX package to subtract a 
linear baseline from each spectrum and average together spectra taken
at the same position.  The averaged spectra were binned in velocity to
a resolution of 3 \kms\ to improve the signal-to-noise ratio.  To convert
the data to the main beam temperature scale, a value of $\eta_{MB}$ = 0.52
from the JCMT User's Guide was adopted.

We compare our [CI] observations with the CO J=$1\to0$ emission
observed by Wilson \& Scoville (1992) using the Owens Valley
Millimeter-Wave Interferometer.  Prior to comparison, the interferometer
data were spatially smoothed to 12\arcsec\ resolution.  Figure~1 shows 
the [CI] spectra, compared with spectra from the same positions in 
the CO J=$1\to0$ line.  

\placefigure{fig1}

\section{Results}

Table~2 lists the central velocities, full width at half maximum (FWHM),
peak temperatures, and integrated intensities for the observed [CI] 
positions, and for the CO J=$1\to0$ line at those same positions. Wilson 
(1997) observed [CI] at 492 GHz in NGC\,604-2 with a single pointing at the 
central position, obtaining an integrated flux of 2.8 $\pm$ 0.4 K \kms.  
Our new measurement of this position agrees very well with the previous 
observation.  Figure~2 shows our spectra superposed on the full resolution 
CO J=$1\to0$ map, and Figure~3 shows the spectra superposed on an H$\alpha$ 
map from the HST archive. 

\tablenum{2}
\placetable{table2}
\label{table2}

\placefigure{fig2}

\placefigure{fig3}

The central velocities of the [CI] line at each location are the same as 
for the CO J=$1\to0$ line, within the uncertainties. Because of the 
agreement in central velocities of the two species, it is likely 
that the [CI] and CO are associated.

Figure~2 shows that the [CI] emission is not distributed symmetrically 
around the peak of the CO emission.  The strongest [CI] emission, 
measured by the integrated intensity I$_{[CI]}$, occurs to the northwest 
of the peak of the CO distribution, on the side of the NGC\,604-2
that faces the center of the giant HII region.  On the opposite side of 
the cloud, [CI] is not detected.

The line ratio I$_{[CI]}$/I$_{CO}$ in NGC\,604-2 varies from 0.11 to 0.23, 
with the mean at 0.18 $\pm$ 0.02.  The line ratio at the (0,0) position is 
consistent with that  obtained by Wilson (1997).  There appears to be a 
gradient in the line ratio in the cloud, in the direction facing the center 
of NGC\,604.  Figure~4 shows the positions observed in [CI] as in Figures~2 
and 3, but with the I$_{[CI]}$/I$_{CO}$ line ratios and N(C)/N(H$_2$) column 
density ratios indicated.  The possible gradient is most evident in the line 
running through the cloud from (0,-5) to (0,10), where the line ratio 
increases towards the northwest edge of the cloud until the end of the cloud 
is reached.  Our observations do not prove the existence of the gradient,
because the difference between positions (0,0) and (0,5) is roughly
the same size as the error bars.  Such a gradient would be expected,
however, due to the higher photodissociation rates at the edge of the cloud.

\placefigure{fig4}

The column density of atomic carbon may be determined from I$_{[CI]}$
under the assumption that the [CI] emission is optically thin:
\begin{equation}
N(C) = 2 \times 10^{15} (e^{23.6/T_{ex}} + 3 + 5e^{-38/T_{ex}}) I_{[CI]}
\end{equation}
\noindent
(Phillips \& Huggins 1981), where T$_{ex}$ is the excitation temperature.  
For T$_{ex}$ we adopt 100 K, which is the kinetic temperature for NGC\,604-2
derived by Wilson et al. (1997) from a large velocity gradient (LVG) analysis 
of CO J=$2\to1$ and J=$3\to2$ lines.   In the above equation N(C) does not 
depend sensitively upon the value of T$_{ex}$; in fact, Wilson (1997) remarked 
that if T$_{ex}$ = 10 K, then N(C) increases by less than a factor of 2.  The 
column densities we obtain range from $2.3~\times~10^{16}$ \cm2 to 
$5.1~\times~10^{16}$ \cm2.

To derive the column density of atomic carbon relative to molecular hydrogen, 
we use the CO J=$1\to0$ data to estimate the column density of H$_2$.  The CO 
column density cannot be derived from these data because the $^{12}$CO 
J=$1\to0$ line is optically thick, but by using the metallicity dependent CO 
to H$_2$ conversion factor of Wilson (1995), we can convert I$_{CO}$ into the 
H$_2$ column density.  The CO to H$_2$ conversion factor is 1.5 $\pm$ 0.3 
times the standard value for the Milky Way Galaxy 
(3~$\pm$~0.3~$\times$~10$^{20}$ \cm2 (K \kms)$^{-1}$; Strong et al.1988).
The H$_2$ column densities vary from 4~$\times~10^{21}$ \cm2 to 
8~$\times~10^{21}$ \cm2, giving column density ratios, N(C)/N(H$_2$), of 
$\sim$ 6 $\times~10^{-6}$,  except at (-5,0), where the ratio is
almost a factor of two less.  

Table~3 lists the line ratio I$_{[CI]}$/I$_{CO}$, the atomic carbon
column density N(C), and the ratio between carbon and molecular
hydrogen column densities, N(C)/N(H$_2$).

\tablenum{3}
\placetable{table3}
\label{table3}

\section{Discussion}

\subsection{Comparison to Galactic Molecular Clouds}

The best comparison to our observations of NGC\,604-2 is the sample of 
four galactic molecular clouds observed by Plume et al. (1999).  They
mapped the clouds in several lines, including [CI] and $^{13}$CO J=$2\to1$.
Like us, they mapped most of each cloud, though they obtained a higher
physical resolution since their clouds are galactic.

\subsubsection{Line Ratios}

Plume et al. obtained line ratios of I$_{[CI]}$/I$_{^{13}CO J=2\to1}$ at 
various positions in their clouds, with an average line ratio of 0.59 $\pm$ 
0.19.  For comparison with our data we convert this to I$_{[CI]}$/I$_{CO}$ 
(where I$_{CO}$ refers to the $^{12}$CO J=$1\to0$ line) using the average 
line ratio for $^{12}$CO J=$2\to1$ to $^{12}$CO J=$1\to0$ 
of 0.7 for the Orion A and B clouds from Sakamoto et al. (1994), and the 
average value of the $^{12}$CO to $^{13}$CO J=$2\to1$ line ratio of 5.5 
$\pm$ 1 measured for the Milky Way by Sanders et al.(1993).
The Plume et al. value of I$_{[CI]} $/I$_{CO}$ becomes 0.08 $\pm$ 0.03, 
slightly lower than would be allowed by the scatter about our mean value, 
but consistent within the error bars with some of our individual positions.

If we adopt different, but still reasonable, values for the two line ratios, 
the results agree with our observations better still.  Using the $^{12}$CO 
J=$2\to1$ to $^{12}$CO J=$1\to0$ line ratio of 1.3 found near HII regions
in Orion instead of the global average, and using the $^{12}$CO to
$^{13}$CO J=$2\to1$ ratio of 4.5 $\pm$ 0.7 found by Wilson, Howe \& Balogh
(1999) for the molecular cloud M17, we find I$_{[CI]}$/I$_{CO}$ = 0.17 
$\pm$ 0.06, in good agreement with our observations.  Clearly because 
Plume et al. did not observe $^{12}$CO J=$1\to0$, comparisons between our 
data and theirs will be inexact.  We do, however, find a general consistency.

On the edges of the molecular clouds, where the CO is less shielded, Plume 
et al. find values of the line
ratio up to 7 times higher.  We see a similar trend in NGC\,604-2, but our
physical resolution at the distance of M33 is 50 pc, compared to the 1 - 2 pc
for Plume et al.  Thus, we are averaging over large areas within the cloud
and smoothing away any large contrasts in the line ratio that may exist.
Observations with a submillimeter interferometer would help in this case.

Plume et al. find a strong correlation between [CI] line strength and
$^{13}$CO J=$2\to1$ line strength, and fit both a power-law and linear
relation to their data.  We can use the $^{13}$CO J=$2\to1$ observation
by Wilson, Walker \& Thornley (1997) to determine if NGC~604-2
follows either relation.  The beam size of Wilson et al. 
is 22$\arcsec$, so we scale their $^{13}$CO J=$2\to1$ intensity by
(22)$^2 / $(12)$^2$ to make an approximate correction for the different
beam sizes.  Using this value in the two fitted relations, we find that
either of them reproduces our I$_{[CI]}$ observation at the (0,0) position.
The linear relation gives I$_{[CI]}$ = 2.7 $\pm$ 0.2, compared to the
measured value of 2.8 $\pm$ 0.3  The power law gives 3.0 $\pm$ 0.2. Our
data point falls close to both relations in the plot of
I$_{[CI]}$/I$_{^{13}CO J=2\to1}$ versus I$_{^{13}CO J=2\to1}$ in Plume
et al., their Figure~10.  Both our data point, and the data of Plume et al.
are in qualitative agreement with single layer PDR models.

\subsubsection{Column Densities}

We obtain column densities a factor of 3 to 7 smaller than the typical
column densities measured by Plume et al. The difference may be explained 
by the lower than solar metallicity of NGC\,604, or the fact that NGC\,604-2 
lies in a giant HII region, compared to the more normal galactic star 
forming regions of the Plume et al. clouds.  They find lower column densities 
at the edges of their clouds, and higher densities at the cloud cores, which 
we do not, but this is again likely due to our lower physical resolution.

We compare the column density ratio N(C)/N(H$_2$) in NGC\,604-2 with 
the observations by Plume et al. (1999) and predictions of models by Flower
et al. (1994) and Spaans \& van Dishoeck (1997).  Plume et al. calculate 
densities of C and $^{13}$CO, and the theoretical papers derive column 
densities of C, $^{12}$CO, and $^{13}$CO.  To convert these CO densities 
to H$_2$ densities, we use a $^{13}$CO/H$_2$ ratio of 1.5~$\times~10^{-6}$ 
(Bachiller \& Cernicharo 1986) and an isotope abundance ratio 
$^{12}$C/$^{13}$C of 62 (Langer \& Penzias 1993).  These numbers come from
molecular clouds relatively close to the solar neighborhood.  As a
consistency check, we will also convert the Plume et al.
$^{13}$CO J= 2$\to$1 data to an H$_2$ column density more indirectly,
using the  $^{12}$CO J=$2\to1$ to $^{12}$CO J=$1\to0$ and $^{12}$CO
to $^{13}$CO J=$2\to1$ line ratios discussed in the last section and
the Galactic CO to H$_2$ conversion factor.  The uncertainties involved
in this indirect determination will be larger than would be the case
simply using the Bachiller \& Cernicharo  density ratio,
but it will provide a rough consistency check.

The average N(C)/N(H$_2$) ratio for the Plume et al. molecular clouds is 
2.4~$\times~10^{-5}$, a factor of four greater than we find for NGC~604-2.  
From I$_{[CI]}$/I$_{CO}$ = 0.17, derived in the previous section, N(C)/N(H$_2$)
is 9~$\times~10^{-6}$, nearly three times lower lower than what we found 
using the Bachiller \& Cernicharo density ratio, and more consistent with
our number for NGC~604-2.  However, as we mentioned, this second method 
is more indirect, assuming that $^{12}$CO J=$2\to1$ to $^{12}$CO J=$1\to0$ 
ratio and the $^{12}$CO to $^{13}$CO J=$2\to1$ ratio from small Milky Way 
molecular clouds is applicable to the much larger NGC~604-2, which is 
located in a giant HII region.  It is reassuring that there is an agreement 
of a factor of three between these two calculations. 

The most obvious differences between the Plume et al. clouds and NGC~604-2 
are that NGC~604-2 is in a giant HII region, rather than a small HII region, 
and that NGC~604-2 is more metal poor than the HII region of Plume et al.
Stronger dissociating radiation could lead to enhanced C creation through
photodissociation.  It is also possible that the lower metallicity of
NGC~604-2 plays a role; with less C, and therefore less CO, the self
shielding of CO against dissociating radiation will be lower, allowing a 
greater fraction of the existing CO to be dissociated.  Of course, lower 
metallicity also means that the overall abundance of carbon will be lower, 
so it is not clear to what degree the enhanced photodissociation 
counters the lower metal abundance.  There is a difference in the physical 
size of the clouds -- the Plume et al. clouds are $\sim$ 10 pc in size, 
compared to $\sim$ 30 pc for NGC~604-2.  Perhaps differing N(C)/N(H$_2$) 
ratios represent a difference in the distribution of gas in the clouds, with
the smaller galactic clouds having lower column density, and hence allowing
greater penetration by dissociating radiation.

The models of Flower et al. give N(C)/N(H$_2$) ranging from $\sim$
10$^{-4}$ to 10$^{-7}$.  They can easily obtain values consistent with
our observations, but the models have at least four free parameters,
including the abundances of carbon and oxygen, the visual extinction,
A$_v$, and degree of ionization.  With observations of just [CI] and CO
we cannot distinguish between their various models.  The models of
Spaans \& van Dishoeck are more constrained because they
attempt to model a specific Galactic molecular cloud, S140.  Their
models assume a clumpy ISM which permits UV radiation to penetrate deep
into the cloud.  They obtain column density ratios approaching ours for
models with low volume filling factors (10\% to 30\%) and with relatively
large clumps (0.4 to 0.6 pc).  Of course, S140 is different from NGC\,604-2;
it is illuminated by a single B0V star, not a giant HII region like
NGC\,604, so the agreement between the model and our observations should
be taken with a grain of salt.
However, lacking a detailed model specifically for NGC\,604-2,
a clumpy ISM would explain our observations.

\subsection{[CI] Observations in Other Local Group Galaxies}

Stark et al. (1997) have measured the [CI] emission towards two regions 
within the Large Magellanic Cloud, N 159-W and 30 Dor N.  In the latter 
region, they obtain a I$_{[CI]}$/I$_{CO}$ line ratio of 0.26, twice the 
average for the Milky Way, but similar to what we have seen in NGC\,604.  
They attribute this large ratio to the lower metallicity of the LMC.  
Bolatto et al. (2000, in press) find a similarly high line ratio, 
0.23 $\pm$ 0.03, in the low metallicity Local Group dwarf irregular galaxy 
IC~10. Our observations are fairly consistent with this 
picture; the metallicity of the NGC\,604 region is nearly half solar 
(Vilchez et al. 1988) while the LMC is about one-third solar (8.37; Garnett 
1990), and IC~10 is one-fourth solar metallicity (Lequeux et al. 1979).

\placefigure{fig5}

Figure~5 shows a plot of I$_{[CI]}$/I$_{CO}$ versus metallicity for 
molecular clouds in Local Group galaxies. The Milky Way Galaxy is also 
included, with the data on I$_{[CI]}$/I$_{CO}$ coming from the COBE data
of Wright et al. (1991). Because COBE did not detect CO J=1$\to$0, we scaled 
the I$_{[CI]}$/I$_{CO J=2\to1}$ ratio by a typical $^{12}$CO J=2$\to$1 / 
$^{12}$CO J=1$\to$0 line ratio of 0.7 (Sakamoto et al. 1994).  Although 
only eight points are plotted, the molecular clouds found near HII regions 
have consistently higher line ratios than those with no nearby HII regions.  
This is to be expected if radiation from HII regions is primarily responsible 
for producing atomic carbon by photodissociation of CO.  There also appears 
to be a trend among the clouds near HII regions for I$_{[CI]}$/I$_{CO}$ to 
decrease with increasing metallicity.  This trend might be expected if the 
column density of CO increases with metallicity, thereby more effectively 
shielding the interiors of the molecular clouds from dissociating radiation.  
It could also be a result of the metallicity dependence of the CO to H$_2$ 
conversion factor -- if I(CI)/N(H$_2$) remains constant, but I(CO)/N(H$_2$) 
decreases with metallicity, a similar effect might be observed.  More data 
are needed to confirm these trends.

\subsection{The Origin of the [CI] Emission}

Figure~2 shows that the spatial distribution of [CI] emission
is offset to the northwest relative to the CO.  This offset is not due 
to the pointing problems which were discussed and corrected in Section~2.  
The center of the giant HII region is northwest of NGC\,604-2.
This is where most of the O stars are found (Drissen et al. 
1993) and hence the source of UV photons.  Thus the 
offset of the [CI] emission is towards the direction from which the 
photodissociating photons are incident, suggesting an edge-on morphology 
for the PDR in NGC\,604-2.   

The fact that [CI] emission is present into the center of the cloud 
indicates a non-negligible ionization fraction in the cloud's interior.
At low ionizations the gas phase chemistry is dominated by reactions
with H$^+_3$, which lead to the conversion of carbon into CO (Graedel
et al. 1982).  One possibility to explain the necessary
ionization inside the cloud is penetration by cosmic rays (e.g. Flower
et al. 1994).  We argue that this mechanism is not likely
to be an important one in the case of NGC\,604-2 based upon the asymmetric 
spatial distribution of the [CI] emission.  For energies of 10$^{18}$ 
eV or less, a galactic magnetic field prevents cosmic rays from escaping a 
galaxy freely (Longair 1994).  These cosmic rays spiral
around the magnetic field lines, lose their original direction
of motion, and become isotropic.  Thus ionization from cosmic rays would
not be expected to produce the asymmetry in [CI] emission that we see.
In addition, Yang et al. (1996) have identified supernovae
driven shells in NGC\,604, and the spatial distribution of these shells
(and hence the supernovae which could provide the cosmic rays) is not
consistent with the [CI] distribution.

We cannot rule out a contribution to the creation of C by cosmic rays, 
but the lack of [CI] emission at the (0,-5) position, where CO emission 
is only 25\% less than the cloud center (see Figure~1), suggests that any 
such contribution is small.  The (0,5) position lies on the opposite 
side of the cloud, but facing the center of the giant HII region, and 
I$_{[CI]}$ here is the strongest we have measured, stronger than at the 
center of the cloud, although the CO emission is nearly the same as at 
(0,-5).  This result suggests that the atomic 
carbon is being created by the photodissociation of CO by photons from 
the massive stars.  The penetration of the photons would then be permitted 
by a clumpy ISM with a moderate or low filling factor (e.g. Stutzki et al. 
1988).  These photons penetrate partly through the cloud, 
even to the center, but do not penetrate all the way through, which 
explains the lack of [CI] emission on the far side.

Churchwell \& Goss (1999) argue that the molecular cloud
lies behind the HII region, based upon a measurement of extinction
towards the cloud that is too low for the molecular mass of the
cloud.  Our observations do not conflict with this idea, as long
as NGC\,604-2 is behind the bulk of the ionized gas along the line of
sight, but still close enough to the HII region to receive the 
influx of UV photons that are responsible for dissociating CO and
thus producing atomic carbon.  Also, if the cloud were too far behind
the HII region, then a position like (0.-5) would not be shadowed by
the cloud center, and thus would show [CI] emission.  

\section{Summary}

We have mapped [CI] emission in a GMC in the
giant HII region NGC\,604, the largest HII region in M33.  The 
spatial distribution of the [CI] emission is offset with respect
to the CO J = 1$\to$0 emission, with the peak of the [CI] emission
sitting in the direction of the center of the HII region.  We have
argued that this reflects the fact that the massive stars in the HII 
region produce an intense UV radiation field which photodissociates CO 
into C and O and provides the ionization necessary for 
the chemical reactions which prevent CO from reforming.  The fact 
that the [CI] emission is seen all the way into the center of the
molecular cloud can be understood in terms of a clumpy ISM with 
a moderate or low filling factor which permits deep penetration of
UV photons into the cloud.  The average line ratio of I$_{[CI]}$/I$_{CO}$
is 0.18 $\pm$ 0.04, and individual values range from 0.11 to 0.23.  Values
of 0.1 are commonly seen in Galactic giant molecular clouds, while 
values of 0.2 are typical for environments of intense star formation
activity.  We find column density ratios N(C)/N(H$_2$) of a few 
$\times$~10$^{-6}$, which are consistent with numerical models of PDRs.

\acknowledgements

We thank Gerald Moriarty-Schieven, Henry Matthews and Lorne Avery for 
performing the remote observing.  We also thank the anonymous referee, 
and the editor, Steve Willner, for helpful comments.  The JCMT is operated 
by the Joint Astronomy Centre on behalf of the Particle Physics and 
Astronomy Research Council of the United Kingdom, the Netherlands 
Organization for Scientific Research, and the National Research Council 
of Canada.  The Five College Radio Astronomy Observatory is operated with 
the permission of the Metropolitan District Commission, Commonwealth of 
Massachusetts, and with the support of the National Science Foundation 
under grant AST-9725951.

\vfill
\eject

\begin{center}
{\bf Figure Captions}
\end{center}

\figcaption[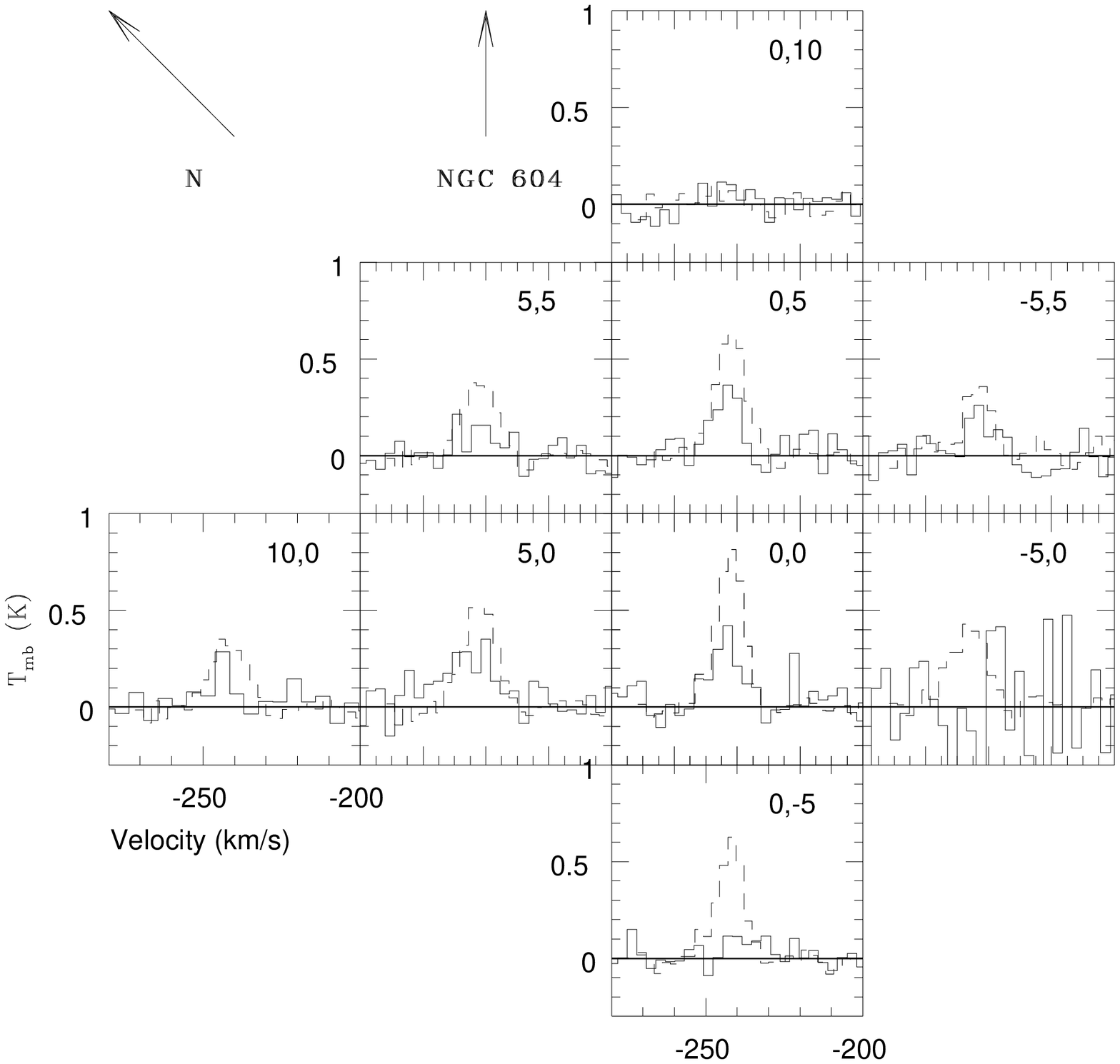] {The [CI] spectra of the positions observed in NGC\,604-2
(solid line).  For comparison, the spectra of CO J = $1 \to 0$ are shown 
(dashed line), divided by 2 to allow plotting with a reasonable scale. The
labels indicate the offsets in arcseconds from the central position (1$^h$
34$^m$ 33.6$^s$, +30$^{\circ}$ 46$^{\prime}$ 48 $^{\prime\prime}$).  
The arrows indicate the directions of north and the center of NGC\,604.
Not shown are the original observations intended for positions (0,10) and 
(-5,0).  Only the re-observed, correct positions are shown.
\label{fig1}}

\figcaption[fig2.ps] {The [CI] spectra from NGC\,604-2 plotted on top of 
a greyscale image of the OVRO map of CO J = $1 \to 0$ emission from Wilson 
\& Scoville (1992).  The spectra are plotted at their true positions, but
labeled with their intended positions for ease of comparison with Figure~1.
\label{fig2}}

\figcaption[fig3.ps] {The [CI] spectra as in Figure~2, plotted on top of an
H$\alpha$ image obtained from the HST archive.
\label{fig3}}

\figcaption[fig4.ps] {The I$_{[CI]}$/I$_{CO}$ line ratio and N(C)/N(H$_2$)
column density ratio (in italics) for the observed positions.
\label{fig4}}

\figcaption[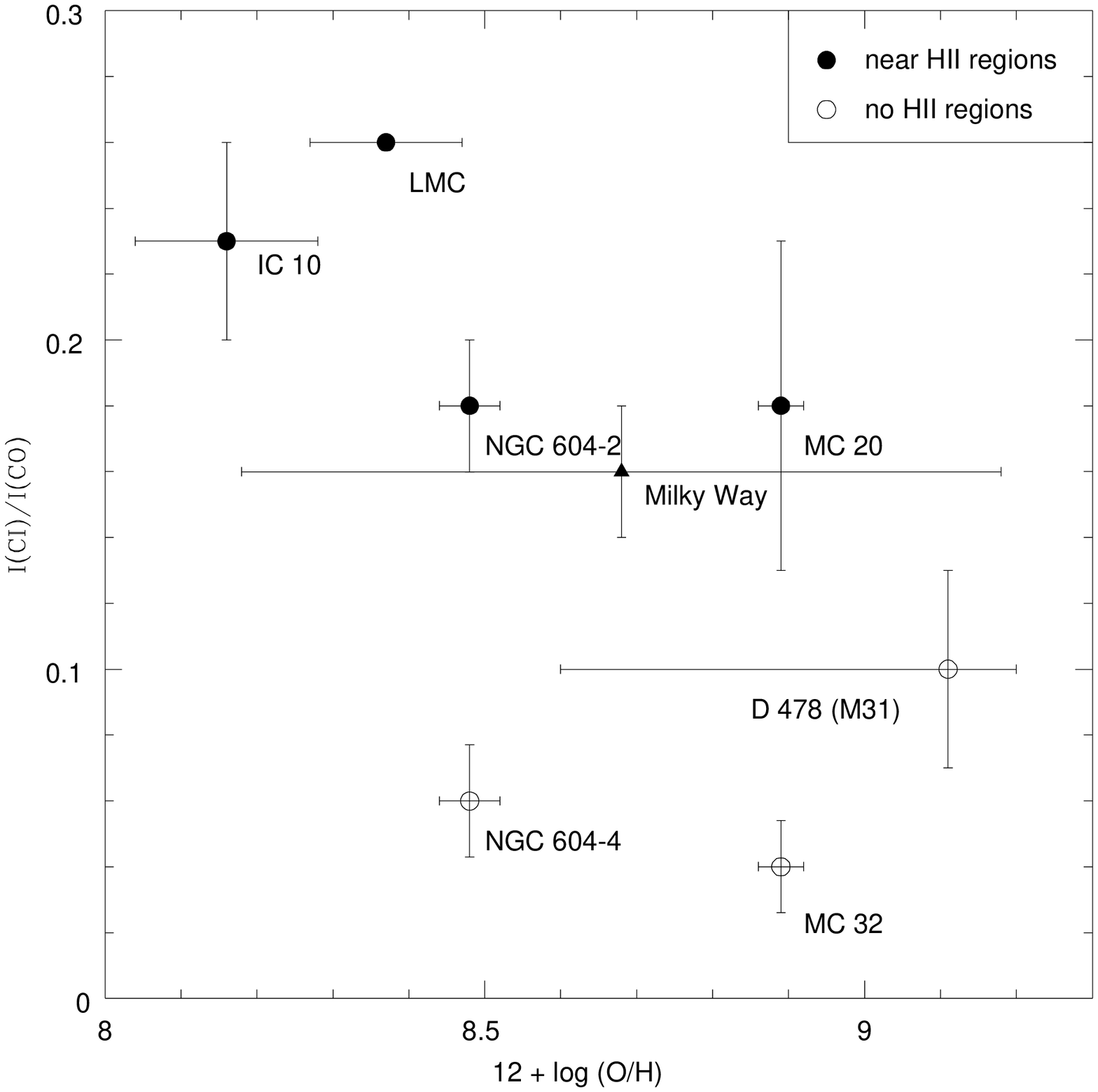] {A plot of I$_{[CI]}$/I$_{CO}$ versus metallicity
for extragalactic molecular clouds in the LMC, IC10, M33 (Wilson 1997) and
M31(Israel, Tilanus \& Baas  1998).  Clouds in the vicinity of HII regions
are plotted as filled circles, and clouds with no nearby HII regions 
present are open circles.  The dark cloud D 478 in M31 is plotted at the
metallicity extrapolated for its location based upon the metallicity gradient
in M31, and the error bars show the range of observed metal abundances.
The average line ratio for the Milky Way galaxy (Wright et al. 1991) is 
plotted at the average metallicity at the solar circle, and the error bar 
indicating the range of metallicity in the Galaxy (Henry \& Worthey 1999). 
\label{fig5}}

\vfill \eject

\end{document}